\documentclass[prd,aps,nofootinbib,nobibnotes,superscriptaddress,twocolumn]{revtex4}
\usepackage{graphicx}
\usepackage[figuresright]{rotating}
\usepackage{amssymb}
\usepackage{bm}
\usepackage{color}

\definecolor{dgreen}{rgb}{0,0.6,0.0}

\newcommand{\al}{\alpha}

\newcommand{\Om}{\Omega}

\newcommand{\be}{\begin{equation}}
\newcommand{\ee}{\end{equation}}

\newcommand{\bea}{\begin{eqnarray}}
\newcommand{\eea}{\end{eqnarray}}
\newcommand{\bean}{\begin{eqnarray*}}
\newcommand{\eean}{\end{eqnarray*}}

\newcommand{\id}{{\rm 1\kern -2.5pt I}}

\newcommand{\eqref}[1]{(\ref{#1})} 

\begin{document}


\title{Catastrophic Dark Matter  Particle Capture}



\author{Ruth Durrer}
\email{ruth.durrer@unige.ch}
\affiliation{D\'epartement de Physique Th\'eorique and Center for Astroparticle Physics,
Universit\'e de Geneve, Quai E. Ansermet 24, 1211 Gen\'eve 4, Switzerland}

\author{Serge Parnovsky}
\email{ parnovsky@knu.ua}
\affiliation{Taras Shevchenko National University of Kyiv, Astronomical observatory, 
Observatorna str. 3, Kyiv 04053, Ukraine}

\date{\today}

\begin{abstract}
In this paper we describe a new idea which may be relevant to the formation of galaxies via the infall of baryonic matter (BM) and dark matter (DM) onto a pre-existing over density. While BM can under certain circumstances be captured by thermal processes, DM particles fly through a static over density without being captured.  We propose a simple model for DM capture: if during the passage through  it, the mass of the over density  increases, then slow DM particles are captured by it, further increasing its mass, while faster particles slow down, transferring part of their energy to the galaxy.
\\
We estimate the minimum initial velocity of a particle required for a passage without capture through the center of the galaxy and derive a nonlinear equation describing the rate of galaxy mass increase. An analysis carried out using the ideas  of catastrophes theory shows that 
if the  increase in the  mass of baryonic matter exceeds a certain threshold value, this can lead to a very
intensive capture of dark matter. We speculate that this process may be associated {on the one hand with the accretion of matter during the early stages of galaxy formation and, on the other hand  also later with the merger of galaxies.} For the studied process to take place, the density of intergalactic DM must exceed some threshold value. Then the rate of increase in the mass of DM can be much higher than the one of baryonic matter.  The capture sharply decreases after the DM density drops below the threshold value, e.g., due to the expansion of the Universe.

\end{abstract}




\maketitle

\section{Introduction}

It is well established that every galaxy is surrounded by a dark halo. This halo consists of dark matter (DM) and its size is much larger than the visible size of the galaxy. The mass of DM halo is the dominant contribution to the total mass of the galaxy.
We  assume that DM consists of some still unknown particles that are not affected by strong and electromagnetic interactions. They interact between themselves and with baryonic matter (BM) only gravitationally.  In the present paper we neglect a possible very weak non-gravitational interaction of dark matter with baryonic matter or with itself .

The formation of galaxies or more generally the large scale structure of the Universe (LSS) is a very important problem in cosmology. The basic picture is that galaxies have grown out of small initial fluctuations from inflation by gravitational instability.  This idea is in good agreement with the small fluctuations observed in the temperature of the cosmic microwave background~\cite{Planck:2018vy,Durrer:2020fza}. As long as perturbations are small, they can be studied with linear or higher order perturbation theory. But for the formation of galaxies, such an approach is not suitable for two reasons. First of all, the over densities of galaxies, $\rho_{\rm gal}/\rho_m \gtrsim 10^5$, are much larger than $1$ so that perturbation theory cannot be expected to converge. Furthermore, we assume DM particles to be collisionless. In this case they should be described by Vlasov's equation in phase space. While this leads to a similar Jeans scale as the fluid approach, just replacing the sound speed by the velocity dispersion~\cite{1966ApJ-144-233G,BinneyTremaine}, non-linear aspects are very different as soon a shell crossing becomes relevant which leads to singularities in the fluid approach.

For this reason, the non-linear regime of cosmological structure formation is usually treated either via N-body simulations, which have, at present, achieved an impressive amount of detail~\cite{Bose:2022ibz,2010MNRAS.407..435A,1991ApJ...367...45C,2004ApJ...616...16G} or by simple analytical models, e.g., spherical collapse or secondary infall models~\cite{1977ApJ...218..592G,1980lssu-book-P,1984ApJ...281....1F,1985ApJS...58...39B,Wang:2019ftp}. 

Even though these works are very important and have provided a rather clear picture of the formation of cosmological LSS, they have their intrinsic limitations: the best N-body simulations have a resolution of $10^{-11}M_\odot$,~(see \cite{Wang:2019ftp}) to $10^{12}M_\odot$, depending on the size of the region they want to simulate and on computational resources. Even $10^{-11}M_\odot$ is much larger than the mass of dark matter particles which have typically masses of elementary particles, $m\sim 100$Gev $\simeq 10^{-45}M_\odot$. This is not quite true in the case of primordial black holes which have a window of possible masses between $10^{-16}M_\odot<m<10^{-11}M_\odot$, but can also be heavier, see~\cite{Carr:2020xqk} for a review. 
In Ref.~\cite{Wang:2019ftp}, the authors have shown that the halo density profiles are universal over
the entire mass range and are well described by simple two-parameter fitting formulae. Nevertheless, at small scales they zoom in  on regions with moderate over densities of at most 17 which is much less than the density of a typical galaxy. Furthermore, these simulations do not include baryonic physics which is crucial for the effect discussed in the present work.

Even though the above results are very important and promising, it is still not entirely clear that mass resolution is irrelevant. Naively, the process needed for the collapse of collisionless particles, dynamical friction, has a cross section which is proportional to $m^2$.
The energy loss of a particle with mass $m$ moving with the velocity $v$ through the media with matter density $\rho$ is given by~ \cite{Longair}
\be\label{e:Edot}
\frac{dE}{dt} = - \frac{4\pi G^2 m^2\rho}{v}\ln\left(\frac{b_{\max}}{b_{\min}}\right),
\ee
where $G$ is the gravitational constant and $b$ is the impact parameter of the collision. Taking into account that $E =(mv^2)/2$, we find that $dv/dt \propto m$.
Might it be that due to the much smaller effect from dynamical friction, elementary particles might behave significantly different from 'chunks of phase space'  with a mass of $10^6M_\odot$ ?

What concerns the analytical models, the problem is that assuming spherical symmetry one can 'circumvent' Liouville's theorem which states that phase space volume is conserved under Hamiltonian evolution.  In a spherically symmetric situation the phase space volume vanishes already in the initial condition since velocities have only $1$ non-zero component. More physically: since all particles move radially, the total angular momentum vanishes and its conservation does not constrain the infall.

In this paper we consider secondary infall into a spherically symmetric over density.  This problem has been addressed before. e.g. in Refs.~\cite{1984ApJ...281....1F,1985ApJS...58...39B}, but only for radially in-falling particles. Here, even though we assume a spherically symmetric gravitational potential, particles may fall in with arbitrary, non-zero impact parameter. {Note, also that we assume DM and BM to interact only gravitationally. There are of course also models where DM and baryons interact with non-gravitational forces, see~\cite{Salucci:2020eqo} for arguments to favor this possibility,  but we neglect such interactions in the present study.}

If the gravitational potential remains constant, the particles will gain  velocity during infall and will loose it again when climbing out of the potential, but they will not be captured.  However, if the potential is growing during the infall, some particles with sufficiently low initial velocities can get captured.

In this paper we show an interesting new phenomenon: if the growth rate of the gravitational potential is sufficient, a catastrophe (in the mathematical sense~\cite{Arnold}) happens, enhancing very significantly the capture rate. This may lead to the formation of heavily DM dominated objects like some dwarf galaxies or low surface brightness galaxies. The astrophysical significance of this new effect for galaxy formation still requires a more detailed study. In this paper we just derive and explain the effect of catastrophic DM capture and we illustrate it with some numerical examples.

Even though this effect may in principle be inside N-body simulations, we present here an entirely new and semi-analytic understanding of very enhanced dark matter capture as it can occur if there is significant baryonic accretion. Initially, N-body simulations of cold dark matter do not include velocity dispersion. However, once shell crossing has occurred, significant velocity dispersion is generated and our process can take place. For warm and hot dark matter, where velocity dispersion is relevant from the beginning, our process can occur as soon a baryons are accreted at a significant rate. For fuzzy dark matter however, which can be considered as a Bose-Einstein condensate, see e.g.~\cite{Harko:2015aya}, our description of DM accretion is not adequate.  

Independent numerical studies will be necessary to investigate the mechanism outlined here in more detail, in order to decide about its relevance for cosmological structure formation. Such simulations will need to capture the quite complicated baryonic physics to determine the accretion rate of baryonic matter. Here we study the total mass accretion as a function of the baryonic matter accretion which we treat as unknown external parameter.

The reminder of the paper is structured as follows: In the next section we discuss the basics of DM capture in a gravitational potential. In Section~\ref{s:cata}, the main section of this article, we show that under certain conditions a fold catastrophe can build up leading to a jump in  the dark matter capture rate. In Section~\ref{Sq} we give some quantitative estimates which show that this may happen during  galaxy formation  and in Section~\ref{s:con} we summarize our findings and conclude.

\section{Capture of DM particles}
\subsection{The capture velocity of DC particles}
The  velocities of DM particles increase as they enter from intergalactic space into the halo of a galaxy and decrease as they leave it. If during the flight the mass of the galaxy is increasing, then slow DM particles are captured by the galaxy further increasing its mass, while faster particles slow down, transferring some of their kinetic energy to the galaxy, reducing its gravitational binding energy. Let us consider this mechanism using a simple model that allows us to draw a number of qualitative conclusions. 

We consider a spherically symmetric matter over density which we call 'a galaxy'. It may represent a dark halo with a visible galaxy inside. The first consists mainly of DM while the second consists mainly of BM. We assume that  the total matter density $\rho$ depends only on the radius, $\rho(r)$. This greatly simplifies the model, but contradicts the results of $N$-body simulations described, e.g., in \S 9.3.3 in \cite{BinneyTremaine}. Spherical symmetry does certainly not apply to BM in spiral galaxies. However, it is reasonable to consider to assume that this simplification does not qualitatively change the behavior we now discuss..
 Within this approximation, a particle moving radially cannot be deflected in any direction. It is clear that the halo does not have a sharp boundary. This does not prevent us from using a reasonable estimate for its radius $R$. We neglect the DM density $\rho(r)$ at $r>R$. This radius which we assume to be constant
 is not a value like $r_{200}$, which changes as galactic halos form (see Chapter 9 of \cite{BinneyTremaine}). We choose $R$ slightly larger than the maximum size of the halo at all stages of the evolution of the galaxy, at which it can be called a galaxy. Taking into account the ambiguous definition of this quantity, we will use various possible values of the halo radius $R$ in numerical estimates, from underestimated to overestimated.

We assume DM particles to be non-relativistic, so that we can use the formulas of Newtonian mechanics. Naturally, nothing prevents us from taking into account the effects of special relativity, but this does not change the qualitative results of our model. We include Hubble expansion to account for the change in the density of dark matter particles in intergalactic space. 

Let's consider a DM particle that approaches the over density with an initial speed $v_0$ at $r\gg  R$. Then at the boundary of the halo the particle velocity is equal to $v(R)=\sqrt{v_0^2+u^2 }$ with $u^2=2GM/R$ and inside halo at a distance $r$ from the center it is equal to
\be
v(r) = \left(v_0^2+u^2+8\pi G\int_r^R\frac{dx}{x^2}\int_0^xy^2\rho(y)dy\right)^{1/2} \,.
\ee
The total mass of the over density is
\be\label{e:Mass}
M=4\pi\int_0^Ry^2\rho(y)dy  \,.
\ee
At the very center the particle velocity reaches a maximum value equal to $\sqrt{v_0^2+\al u^2}$. The factor $\al$ is equal $1.5$ for a constant density of matter inside the galaxy. It is easy to calculate it for any given density distribution, e.g. for Navarro-Frenk-White profile~\citep{NFW}. However, for a reasonable density distribution in which the density decreases with distance from the center, the value of $\al$ is not much larger than $1.5$ and,  for a rough estimate, we may set $\al\simeq 1$ and $v(r)\simeq v(R)$ 
at $ r<R$.

If the gravitational field is static,  the particle will leave the halo again with speed $v(R)$. Far away from the galaxy it will move again with speed $v_0$. There is no particle capture. However, capture is possible if the mass of the galaxy  increases during the passage of the particle. Indeed, galaxies continue to grow also after their formation. This can happen by the accretion of the surrounding BM, by the capture of DM particles, and also by merging with other galaxies. 
As a result, the gravitational potential well formed by the galaxy becomes deeper, and the potential barrier surrounding it becomes higher. The DM particle may not have sufficient kinetic energy to leave the potential well and it can be captured. 

We denote by $M$ the total mass of the galaxy at the moment the DM particle enters  ($|x_p|=R$), and by $\tau$ the time of flight of the particle through the galaxy. Then, at the time the particle leaves the galaxy, the mass of the galaxy will be equal to $M+\dot M\tau$, where $\dot M$ is the average rate of mass increase of the galaxy. The particle capture condition takes the form
\be\label{e:capt}
v_0\leq \sqrt{\frac{2G\dot M\tau}{R}} \,.
\ee
It is reasonable to expect that the mass increase $\dot M\tau$ does not exceed the mass $M$. Therefore, the maximal initial speed of captured particles is less than $u$ and we can estimate $v(R)\simeq u$ so that
\be\label{e:tau}
\tau \simeq \frac{\ell}{u}
\ee
for the time of flight of particles with a minimum initial velocity at which they are not captured by the galaxy. Here $\ell$ is the length of the path traveled inside the halo.

It may seem that these approximations are too crude, but they are not. We demonstrate this with an example. As is well known, the rotation curves of  galaxies are perfectly flat if the density decreases like $\rho(r)=M/(4\pi Rr^2)$ at $r<R$. In this extreme case, the density diverges at the center, and this density profile is usually modified at small radius to avoid this divergence. Let us, however, consider the unmodified profile. In the framework of classical mechanics, the velocity of  a particle passing through the center  is then equal to
  $v(r)=u\sqrt{\ln\frac{R}{r}}$ at $r<R$ if $v_0\ll u$. (The factor
$\al$ introduced below eq.~\eqref{e:Mass} is infinite in this case.)  But we are interested in the time of flight, which is equal to $\tau=2CR/u$ with  $C =e\sqrt{\pi}  {\rm erfc}(1)\simeq 0.76$. So, the estimate \eqref{e:tau} deviates from the exact value of $\tau$ only by 25\% even for this extreme density profile.

Let us denote the minimal initial velocity of a DM particle which is able to fly through the galaxy and escape from it by $v_p$ (the subscript $p$ indicates passage). For $v_0<v_p$ the particle is captured by galaxy. If a particle flies through the center we denote its minimal initial velocity by $ v_{pc}$ (the subscript $pc$  indicates passage through the center). Within the above approximations we obtain the  estimates
\bea
v_p &=&\sqrt{\frac{G\dot M\ell}{Ru}}=\left(\frac{G\dot M^2\ell^2}{2RM}\right)^{1/4}\,, \label{e:vp}\\
v_{pc} &=& \sqrt{\frac{2G\dot M}{u}}=\left(\frac{2G\dot M^2R}{M}\right)^{1/4}\,. \label{e:vpc}
\eea

We can use the simplest estimate $\dot M\simeq M/T,$ where $T$ is the age of the galaxy and find with \eqref{e:vpc}
\be
v_{pc}\simeq \sqrt{\frac{2GM\tau}{RT}} =u\sqrt{\frac{\tau}{T}} \,.	\label{e:tt}				
\ee

\subsection{Evolution of the DM particle velocity}

If a particle escapes from the galaxy, its velocity far from it, $v_1$, is smaller than the initial velocity $v_0$ due to the growth of the galaxy mass,
\be
v_1^2=v_0^2-\frac{2G\dot M\tau}{R}  \,.					\ee
Each particle reduces its speed as it passes through a growing galaxy, transferring the released kinetic energy to the DM and BM inside the galaxies. This deceleration mechanism is similar to the integrated Sachs–Wolfe effect. It works more efficiently in galaxy clusters and poorly in voids, simply because of the difference in the number of galaxies that a particle passes through in the same amount of time. Therefore, we expect the mean kinetic energy of DM  particles  to be higher in voids than in superclusters.

For fast particles with $v_0\gg u$ we can set $v(R)\simeq v_0$. If such  particles fly a path of length $\ell$ inside the galaxy, then $\tau\simeq\ell/v_0$ and
\be
v_1\simeq\sqrt{v_0^2-\frac{2G\dot M\ell}{Rv_0 }}\simeq v_0-\frac{G\dot M\ell}{Rv^2_0} \,.			\ee
The faster the particle, the less is the loss of speed. Therefore, the initial velocity distribution of particles not only shifts in the direction of decreasing velocities, but this shifts depends on $v_0$, changing its velocity spectrum.

Let us estimate the rate of energy loss for a DM particle with mass $m$. During the time $\tau $ of passage through the galaxy, it transfers the energy $\frac{Gm\dot M\tau}{R}$  to it. In this case, the distance traveled is  $\ell\simeq v(R)\tau$. The rate of energy loss per unit time and per unit path are
\be
\dot E =\frac{Gm\dot M}{R}\, , \qquad     \frac{dE}{d\ell}\simeq \frac{Gm\dot M}{Rv(R)}\,.		\ee
The trapped particles transfer all their kinetic energy and their mass to the galaxy. The fact that the motion of the particles  in the non-stationary  field of the contracting matter is an efficient mechanism for the dissipation their energy has also been mentioned in the past in connection with other problems. Consider, for example, Ref.~\cite{1980SvJNP..31..664Z}.

\section{Catastrophic DM capture}\label{s:cata}

\subsection{The rate of increase in the galaxy mass}

We have already mentioned various mechanisms for increasing the mass of galaxies. Let us denote the rate of galaxy mass increase due to  DM particle capture by $\dot M_{DM}$, the total rate of galaxy mass increase due to all effects by $\dot M$, and the rate of mass increase due to accretion of baryonic matter by $\dot M_b$.  Obviously 
\be\label{e:Mdotsum}
\dot M=\dot M_{DM}+\dot M_b \,.
\ee	
We consider $\dot M_b$ as a given, external quantity that can change with time and is determined in part by non-gravitational processes of cooling which we do not investigate here. Even though a part of the baryonic matter, e.g. stars, does behave like collisionless particles, we assume that some baryonic matter, like e.g. gas, is accreted via collisional processes and by emitting radiation. We do not want to study this in any detail but consider the baryon accretion rate as an external parameter $\dot M_b$. We now study  $\dot M$  as a function of the baryon accretion rate, $\dot M_b$. 

We consider particles in extragalactic space far from galaxies. We assume that their velocities are distributed isotropically in the reference frame of the galaxy, and the number of particles with velocities in the range from $v_0$ to $v_0+dv_0$ in a unit volume is equal to $dN=f(v_0 )dv_0$. The total density of DM particles in extragalactic space is $N=\int_0^\infty f(v_0 )dv_0$. Then the number of particles flying in extragalactic space through an area dS into a solid angle $d\Om$   in a time $dt$ with velocities in the range from $v_0$ to $v_0+dv_0$ is
\be\label{e:dN}
dn=\frac{v_0}{4\pi} \cos(\phi)    f(v_0 )dv_0 dSd\Om  dt \,,			
\ee
where $\phi$    is the angle between the particle velocity direction and the normal to the area.

Consider a sphere of radius $R_1\gg  R$, surrounding the galaxy and concentric to it. Its surface area is $4\pi R_1^2$. The number of particles passing through it in a time $dt$ with velocities in the range from $v_0$ to $v_0+dv$ is given by equation~\eqref{e:dN}. The halo is reached by particles emitted into a solid angle $\Om  \simeq \pi R^2/R_1^2\ll   1$. Let us define the function  $k(v_0,\phi )$ which is equal to $1$ if a   particle with angle $\phi$ and velocity $v_0$ is trapped and $0$ if it is not trapped.

With this we obtain for the rate of increase in the mass of the galaxy by DM particles capture
\be
\dot M_{DM}\simeq 2\pi R_1^2m\int dv_0\int d\phi 
\sin\phi\cos\phi f(v_0)v_0 
k(v_0,\phi)\,.
\ee		
Let us evaluate this integral. Considering that our model is rather a toy-model, not very accurate estimates are applicable. In order for a particle to be captured by a galaxy, it must enter it. This happens  if $0\leq\phi   \leq\arcsin(\beta ) $ with $\beta =R/R_1\ll   1$. We therefore can set $\sin\phi    \simeq\phi$   and  $\cos\phi \simeq  1$. Let us also introduce the variable   $\xi=\phi   /\beta$ . If we neglect the curvature of the particle trajectory inside the galaxy (this does not significantly affect the path length for particles flying through), then the path length inside the galaxy is about  
\be
\ell=2R\sqrt{1-\phi^2/\beta^2} =2R\sqrt{1-\xi^2} \,. \ee
Capture occurs and $k= 1$ if the condition \eqref{e:capt} is met, that is, if
\be
1-\xi^2>\frac{v_0^4 (v_0^2+u^2 )}{4(G\dot M )^2}\simeq \frac{v_0^4 u^2}{4(G\dot M )^2} = \left(\frac{v_0}{v_{pc}}\right)^4 \, .			\ee
(Remember that the initial velocity of captured particles, $v_0$,  is much smaller than the escape velocity $u$).
The particle is captured if the variable $\xi$, which is proportional to the angle of deviation of the particle velocity from the center of the halo $\phi$, does not exceed the value
\be
\xi_0^2 (v_0 )\simeq \max\left(0,1- \left(\frac{v_0}{v_{pc}}\right)^4\right) \,.
\ee		
A particle flying through the very center of halo is captured if its initial velocity is less than $ v_{pc}$ given in \eqref{e:vpc}. In an off-center passage, the particle is captured if
\be
v_0\leq v_p=v_{pc} (1-\xi^2)^{1/4}.			
\ee
If this inequality is not satisfied, then there is no capture and k=0. With this we can write
\bea
\dot M_{DM} &\simeq& 2\pi R^2m \int_0^{\infty} f(v_0 ) v_0dv_0 \int_0^{\xi_0} k(v_0,\xi)\xi d\xi \nonumber \\
&=& \pi R^2m\int_0^{\infty} \xi_0^2  f(v_0 ) v_0 dv_0 \nonumber \\
&=& \pi R^2m\int_0^{v_{pc}}\left(1-\frac{v_0^4}{v_{pc}^4} \right)  f(v_0 ) v_0 dv_0 \,.	
\label{e:Mdmdot}		
\eea		
The integral on the right hand side depends on $\dot M$ via $v_{pc}$, and this dependence is highly non-linear. The combination of \eqref{e:Mdotsum} and \eqref{e:Mdmdot} determines $\dot M$ for a given $\dot M_b$ and a given velocity distribution $f(v_0)$. At $\dot M_b=0$ there is a trivial solution $\dot M=0$.

Assuming the form of the function $f$, we can obtain the dependence of $\dot M_{DM}$ on $\dot M$. For example, if  $f$  is a simple Maxwell-Boltzmann  distribution,
\bea
f(v) &= &\frac{4Nv^2}{\sqrt\pi v_{max}^{3}}\exp{\left[-\left(\frac{v}{v_{max}}\right)^2\right]}, \\
 v_{max} &=&\sqrt{\frac{2k\Theta}{m}}  \nonumber
\eea
with temperature $\Theta$ and maximum at $v=v_{max}$, then from \eqref{e:Mdmdot} one obtains
\be
\dot M_{DM}=2\sqrt{\pi} R^2mNv_{max}P(v_{pc}^2/v_{max}^2)\,, \label{e:Max1}
\ee
where the function $P$ is given by
\be
P(x)=1-6x^{-2}+2e^{-x}(1+3x^{-1}+3x^{-2}).\label{e:Max2}
\ee

Of course, we cannot really assume that DM particles obey a thermal distribution. However, interesting qualitative conclusions can be drawn from general considerations without detailed assumptions about the velocity distribution of the DM particles (more precisely, their phase space number density  $f$ or mass density $mf$). We just assume that the function $f(v_0 )\geq 0$, that is it continuous, that it vanishes at $v_0=0$, reaches a maximum at some value $v_0=v_{\max}$, and quickly decreases at high velocities, most likely exponentially in $v_0^2$. This function is proportional to the particle density $N$.

We consider the expansion of $f$ in a Taylor series. It includes only even powers of $v_0$. As for the Maxwell distribution, the expansion starts with a quadratic term due to the three independent Cartesian velocity components:
\be
f(v_0 )=\sum_{i=1}^\infty a_i v_0^{2i} = N\sum_{i=1}^\infty \tilde a_i v_0^{2i} \, .		
\ee
The quantities $\tilde a_i= a_i/N$ do not depend on $N$. So that
\bea
\dot M _{DM} &\simeq& \pi mR^2\int_0^{v_{pc}}\left(1-\frac{v_0^4}{v_{pc}^4}\right) \sum_{i=1}^\infty a_i v_0^{2i+1} dv_0   \qquad 
\nonumber \\
&=& \pi mR^2 \sum_{i=1}^\infty \frac{a_i}{(1+i)(3+i)}v_{pc}^{2i+2} \nonumber \\
& =&\dot M \sum_{i=1}^\infty b_i \dot M ^i =\dot M N\sum_{i=1}^\infty \tilde b_i \dot M ^i  \,\quad\mbox{with} \label{e:MdotMdmdot}\\
b_i &=& \pi mR^2\left(\frac{2GR}{M}\right)^{(i+1)/2}\frac{a_i }{(1+i)(3+i)}\,. 
\eea	 
Also here we have introduced  $\tilde b_i=b_i/N$. These quantities help to explicitly extract the dependence on the particle density $N$, which varies significantly over the lifetime of galaxies. If DM particles are not be formed and do not decay, then $N(z)=N(0)(1+z)^3$ and during the existence of a galaxy formed at $z=10$, the density  decreases by a factor 1000 
due to the expansion of the Universe. However, due to the considered capture of DM particles, their number density in intergalactic space, $N$ is reduced even further.

The quantities $\tilde a_i$ change with time due to changes in the velocity distribution, in particular, because of the processes under consideration. However, most likely, the most significant contribution to the  time dependence of the parameters $a_i$  is associated to the evolution of $N$. The same can be expected for the coefficients $b_i$, although in this case the (weak) dependence of $\tilde b_i$ on time gets additional contributions from the evolution of $M$.

Positivity of $f$ for small velocities requires that $b_1>0$. We assume that $b_2<0$ like for the Maxwell-Boltzmann distribution. 

Let us try to draw some conclusions based on these basic properties of the distribution function. We start with the case when captured DM particles have initial velocities much smaller than their mean velocity. So that they are described by the first term of the expansion of $f(v_0)$,
\be\label{e:nfirst}
f(v_0 )\simeq a_1 v_0^2  \,. 		
\ee
 Let us also assume that the considered galaxy is not very large and the time of passage of particles through it is much less than its age and than the age of the Universe. Then we can approximately set $a_1=$ const. and $M=$ const. during the time of flight.
In this case
\bea
\dot M -\dot M _b &=& \dot M _{DM}\simeq  b_1 \dot M^2=\frac{\pi a_1 R^3 mG}{4M} \dot M ^2 \,,   \nonumber\\
\dot M _b &=& \dot M -b_1 \dot M ^2=\frac{1}{4b_1}  -b_1\left(\frac{1}{2b_1} -\dot M  \right)^2.	 \qquad \nonumber\\
&&\label{e:MDM2}
\eea
Note that $a_1$ has the dimensions $[(\ell^2/t)^{-3}]$ so that $b_1$ has the dimensions $[t/m]$ and $b_1\dot M$ is dimensionless.
For $\dot M _b=0$ this gives us the equation
\be
\dot M \left(1-\frac{\pi}{ 4} a_1 R^3 mG \frac{\dot M}{ M}\right)=0 \,.	\label{e:Mdot2}	
\ee
Only the trivial solution, $\dot M=0$ is physical.
Requiring that the second factor vanishes gives us a rough estimate of characteristic time of mass accretion, $T=M/\dot M$, as\\  $T\simeq  \pi N \tilde a_1 GR^3 m/4$ in\eqref{e:tt}. This is clearly  an unphysical solution with $T \propto N$.  So that, $T$ tends to $0$ with $N=0$,
or, in other words, $\dot M_{DM}$ grows indefinite when $N$ tends to $0$, i.e., when DM becomes less and less abundant, which is meaningless and is a consequence of our approximation which breaks down when $\dot M$ becomes large.

\subsection{A jump in the particle capture rate}
\begin{figure}\begin{center}
\includegraphics[width=8cm]{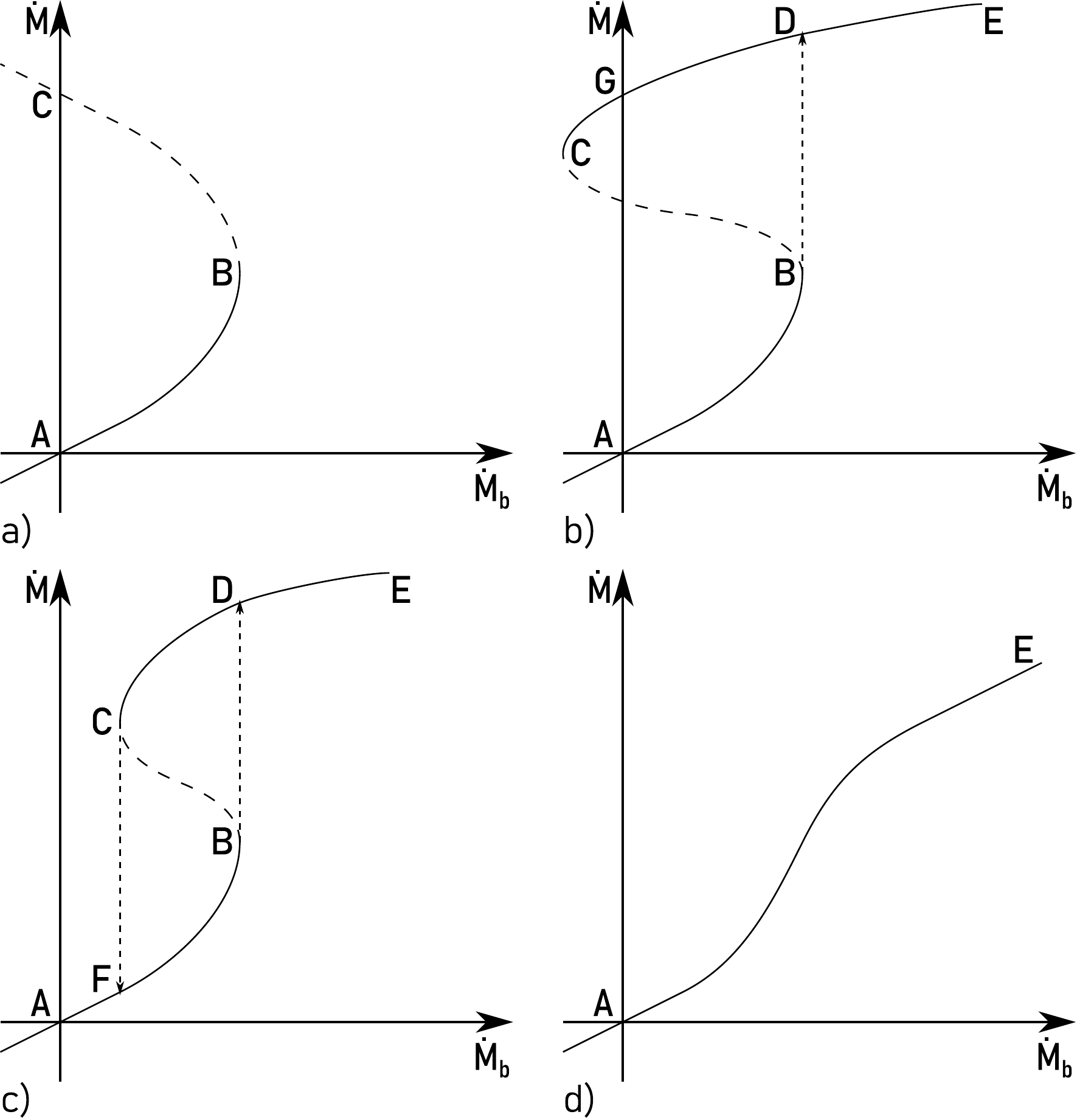}
\end{center}
\caption{\label{f:1}Plots of $\dot M(\dot M_b)$ for various functions $f$ and DM densities. Solid curves show stable branches, dashed curves show unstable ones, vertical dotted arrows show jumps in the state of the system. The four panels correspond to the following cases. Panel a) depicts approximation \eqref{e:nfirst} for low particle velocities. Panels b) and c) show two possibilities for density above threshold. Both are s-shaped. They differ in the position of the left boundary of the upper stable branch at point C. Panel d) shows the case when the baryon accretion rate is below the threshold $ \dot M_{bc}$ defined in Eq.~\eqref{e:Mbc} and the function becomes monotonically increasing.}
\end{figure}
In order to understand this situation, we apply the methods of catastrophe theory. For this it is important to note that we consider $\dot M_b$ as an external parameter which is determined by accretion and baryonic cooling processes which we do not describe in our model. We want to study the increasing of total mass for a given $\dot M_b$. 

Fig. \ref{f:1}a shows the dependence $\dot M$ on $\dot M_b$ according to equation \eqref{e:MDM2}. This fold consists of two parts. The lower half of the parabola AB (in solid) corresponds to a stable solution. With an increase in the baryon mass growth rate $\dot M_b$, the DM particle capture rate $\dot M_{DM}$ increases. It is described by \eqref{e:MDM2} The upper half of the parabola BC (dashed) with a negative slope corresponds to an unstable solution. Two points A and C of intersection of the curve with the y-axis correspond to two the solutions of the equation \eqref{e:Mdot2}. Of these, only the solution $\dot M=0$ is stable. 

At a nonzero matter accretion rate $\dot M_b$, particle capture begins. However, the rate of increase in the mass of the dark halo
$\dot M_{DM}$ is less than the rate of increase in baryonic matter $\dot M_b$. They become equal only at the top of the parabola. At point B we have 
\be\label{e:Mbc}
\dot M_b=\dot M_{DM} =1/(4b_1)\equiv \dot M_{bc}\,.
\ee
From this, two conclusions can be drawn. First, within the approximations made in our model, the capture of dark matter requires the accretion of ordinary matter. There is no DM capture without the baryonic matter  accretion. Secondly, at the ratio of the mass growth of baryonic and dark matter described by the curve in Fig. 1a, it is not possible to form a galaxy containing 85\% dark matter. 

The system~\eqref{e:MDM2} does not have a solution for $\dot M_b>1/(4b_1)$. 
With a further $\dot M_b$ increase, the state of the system reaches the top of the parabola, after which a sudden regime change begins.  and the rate of mass growth $\dot M$ and $v_{pc}$ rapidly increases 
Therefore, we cannot consider the particle velocities to be small and use the approximation \eqref{e:nfirst}. Equation \eqref{e:MDM2} ceases to adequately describe the process. The accretion rate 
$\dot M_b= 1/(4N\tilde b_1)$, which is required to loose stability in point B, can be quite small at the time of galaxy formation, when $N$ is very large.

To consider larger accretion rates we take into account the next term in the expansion  \eqref{e:MdotMdmdot} which leads to  the equation
\be\label{e:MDMdot2}
\dot M_{DM}=\dot M-\dot M_b \simeq b_1 \dot M^2+b_2 \dot M^3,\quad b_2<0.	\ee

Figures 1b and 1c  show the curves ABCDE which can be obtained in this case. They have a characteristic s-shaped form, typical for the fold catastrophe which is  well known in catastrophe theory, see e.g.~\cite{Arnold}. It consists of three parts, two of which have a positive slope (AB and CDE) and are stable. BC with a negative slope is unstable. When the end of the stable part is reached, a jump from B to D to the second, upper stable branch occurs. The upper stable solution can lead to very significant DM accretion, since for it the ratio of the mass growth rates of DM and baryonic matter can be quite large. We call this branch the regime of {\it catastrophic DM capture}. In order to enter this regime, the baryon accretion rate must exceed the
threshold value $\dot M_{bc}\simeq 1/(4N\tilde b_1)$ corresponding to a jump in the capture rate $\dot M_{DM}$.

Note also that a kind of a hysteresis loop can occur in the situation shown in Figure 1c: if $\dot M_b$ is decreasing during the regime of catastrophic DM capture (upper branch), the state of the system on the graph shifts to the left along the upper stable branch. When it reaches the edge of the stable branch at point C, it jumps back  to the lower branch at point F and leaves the regime of catastrophic DM capture. With a certain ratio of the coefficients $b_1$ and $b_2$ this formally happens at a negative value of $\dot M_b$ as is the case in panel 1b. It is easy to calculate, that point C corresponds to a negative value of $\dot M_b$ if $N>-4\tilde b_2 \tilde b_1^{-2}$ 
and a positive value of $\dot M_b$ for  $N<-4\tilde b_2 \tilde b_1^{-2}$. 

Thus, depending on the evolution of $f(v_0)$, the value of $\dot M_{DM}$ can remain in the catastrophic DM capture regime if in the past the galaxy had a value of $\dot M_b$ larger than the critical value and there was a jump, even if later $\dot M_b$  decreases or vanishes. As the particle density $N$ decreases, the left boundary of the upper stable branch, i.e. the point C, crosses the y-axis and the system can return to the lower stable branch via the CF transition.

It is clear that we cannot in general restrict ourselves to a finite number of expansion terms in 
\eqref{e:MdotMdmdot}.  Therefore we now study the  qualitative form of the dependence of
$\dot M$  on $\dot M_b$ without using the series expansion. We introduce the new variable $\eta=v_0/v_{pc}$. With \eqref{e:Mdmdot} 
we obtain
\bea 
\hspace*{-5mm}\dot M_{DM}\!\!\! &\simeq& \pi R^2 mv_{pc}^2F,\\ F \! &=&\int_0^1(1-\eta^4 )  f(\eta v_{pc} )\eta d\eta \nonumber \\
&=&\int_0^1(1-\eta^4 )  f\left( \frac{\eta}{\eta_0} v_{max} \right) \eta d\eta \,, \label{e:26}  \\
\eta_0 &=&\frac{v_{max}}{v_{pc}}.
\eea
The integral $F$ goes over a fixed interval. The integrand is the product of the function $\eta(1-\eta^4)$ that vanishes at both ends of the interval and an unknown function $f$ whose properties we 
discussed above. The function $f$ reaches its maximum at $v_0=v_{\max}$ , i.e., at $\eta=\eta_0$. For small $v_{pc}$ we have $\eta_0\gg 1$. The maximum of $f$ lies outside the region of integration and the integral is proportional to $\eta_0^{-2}\propto v_{pc}^2$. As a result, we can approximate $f$ by
 \eqref{e:nfirst}. 
At large $v_{pc}$ we have $\eta_0\ll 1$ and the maximum of $f$ shifts to the lower boundary of the interval. The integral is approximately proportional to $\eta_0^{2}\propto v_{pc}^{-2}$, which is compensated by the pre-factor $v_{pc}^2$.  Note that here 'large' and 'small'  $v_{pc}$ is considered wrt $v_{\max}$. Hence the colder the DM, i.e. the smaller $v_{\max}$, the less baryonic matter accretion is required to be in the 'large' $v_{pc}$ regime.

A more accurate estimate for the large $v_{pc}$ regime can be obtained directly from the expression \eqref{e:Mdmdot}, in which the upper limit of integration is replaced by infinity, which gives a negligible error in this case. As a result, we obtain the asymptotic expression
\bea
\dot M_{DM} &\simeq&  \pi R^2\int_0^\infty\left(1-\frac{v_0^4}{v_{pc}^4}\right) mf(v_0 ) v_0 dv_0 \nonumber \\
&\simeq&C_1-C_2 v_{pc}^{-4}
=C_1-C_3 \dot M^{-2},   	\label{e:27}	 \\		
C_1&=& \pi mR^2\int_0^\infty f(v_0 ) v_0 dv_0 >0\, , \label{e:28}\\
 C_1 &\propto&  N \, ,  \nonumber  \\
C_2&=&\pi mR^2 \int_0^\infty f(v_0 ) v_0^5 dv_0 >0,\,  \\	
C_2 &\propto& N \,   ,\\
C_3 &=& \frac{M}{2GR}C_2 \,.
\eea
Approximating $C_2\sim C_1v_{\max}^4$, we find that at  high accretion rate, $v_{pc}\gg v_{\max}$ we may neglect the second term in \eqref{e:27} and the DM capture rate is saturated. The dependence acquires the asymptotic form $\dot M\rightarrow C_1+\dot M_b$. In Figures 1b,c, the slopes of the curves in the upper right corner are not drawn to scale.

Let us also consider the behavior of the function for intermediate values of $\eta_0$. The integral $F$ in \eqref{e:26} is a function of the variable $v_{pc}$, that decreases at large and small values of the argument. Therefore, it reaches a maximum at a certain value of $v_{pc}$ which we call $v_{pcm}$. We have $\frac{\partial F}{\partial v_{pc}}=0$ for  $v_{pcm}$. The value of $F$ is proportional to $f(v_{pcm})$, which, in turn, is proportional to the density of particles in intergalactic space $N$. The integral $F$ is multiplied by the factor $v_{pc}^2\propto\dot M$. Therefore, for  $v_{pc}=v_{pcm}$ we obtain a dependence of the form
\be
\dot M_{DM}\simeq  Q(M,R)mN\dot M  			\label{e:29b}		
\ee
where the function $Q(M,R)$ does not depend on $\dot M$.
On the other hand, $\dot M_b=\dot M-\dot M_{DM}$. For small $\dot M$ we have \eqref{e:MDM2} with  positive slope. For large $\dot M$ we have~\eqref{e:27} also with  positive slope. At the maximum of $F$ we have \eqref{e:29b} with the slope $1-Q(M,R)mN$, which is negative if the DM particle density $N$ exceeds some critical value. If this happens, we obtain an s-shaped curve 
$\dot M(\dot M_b)$  like in Fig. 1b and 1c. We apply the theory of catastrophes and find that with a continuous increase of $\dot M_b$, a jump in $\dot M$ occurs and something like a hysteresis loop can appear. This shows that the appearance of the fold catastrophe is quite generic.

It is clear that the catastrophic capture regime must lie above the point with a negative derivative with respect to $\dot M_b$, which is attained at   $v_{pcm}$ for which $F$ is maximal. Let us evaluate this maximum. The integrand in $F$ is the product of two functions, each of which has a maximum. The function $f$ reaches its maximum at $v_0=v_{max}$ , i.e. at $\eta=\eta_0$. The maximum of the function $\eta (1-\eta^4)$ achieved at $\eta=\eta_1\approx 0.7$. The integral is maximal if these two maxima roughly agree,  hence $\eta_0\approx\eta_1$. In this case, the speed $v_{pc}$ for the upper stable branch is about $v_{pcm}\simeq 1.4v_{\max}$. A DM particle with an initial velocity $v_0=v_{max}$ is then captured by the galaxy if its trajectory passes at a distance less than $0.85R$ from the center. This means that a significant fraction of the DM particles is captured as they pass through the galaxy.

The ratio of the influx rates of dark and baryonic matter can be quite large. But the jump into the catastrophic capture regime is not possible if the rate of accretion of baryonic matter did not exceed some threshold value in the past or present. 

The jump also requires the presence of DM with a density $N$ exceeding a certain threshold value. Taking into account that $N$ decreases with time both due to Hubble expansion and because of the capture of particles as discussed in this paper, it can be assumed that eventually the s-shaped curve has turned or will turn into the monotonic dependence  shown in Fig. 1d, where no catastrophe exists and the capture process is significantly  weaker. This moment may lay in the past or in the future depending on the parameters of a given galaxy.  With $\dot M$ also   $v_{pc}$ decreases significantly.

Let us also determine the positions of the points B and C in figures 1b,c which determine the baryon accretion rate at the entry into and the exit from the catastrophic DM capture regime. They are given by the condition $d\dot M_b/d\dot M=0$. Taking into account \eqref{e:Mdotsum} this yields $d\dot M_{DM}/d\dot M=1$. With \eqref{e:Mdmdot} we can write this condition as
\bea
H(v_{pc}) := N^{-1}v_{pc}^{-6}\int_0^{v_{pc}}v_0^5f(v_0)dv_0
 \hspace{2cm}\nonumber\\=\frac{1}{2\pi R^2mN}\left(\frac{M}{2GR}\right)^{1/2}.\label{v1}
\eea
The function $H(v_{pc})$ tends to zero as $v_{pc}\to 0$ and as $v_{pc}\to \infty$. This means that it has a maximum at a certain $v_{pc}=v_1$. It can be estimated that $v_1\approx v_{max}$. We denote
\be
N_1=\frac{1}{2\pi R^2mH(v_1)}\left(\frac{M}{2GR}\right)^{1/2}.\label{N1}
\ee
At $N<N_1$ we have no solution and the inflection points B and C do not exist. The value $N=N_1$ corresponds to the transition from s-shaped curve to the monotonic one. Assuming the generic shape of monotonic increase and decay for $H(v_{pc})$, at $N>N_1$ we have two solutions of  equation \eqref{v1}. The solution with smaller $v_{pc}$ corresponds to point B, one with larger $v_{pc}$ to point C. For small $v_{pc}$ we can use the approximation \eqref{e:nfirst} and obtain the same estimate for coordinates of point B. 
\be
\dot M_b = \dot M_{bc} \simeq  \frac{1}{4N\tilde b_1} =\frac{M}{\pi NR^3 mG\tilde a_1}.\label{e30}	 		
\ee

Using the function $H$, we can show that more complex scenarios are possible in which the transition to the state of intense capture and/or exit from it can occur in two stages. This is possible, e.g., in the case of the existence of two different types of DM particles or a bimodal velocity distribution $f(v)$  \cite{2022arXiv220901819D}.

The point C corresponds to  $\dot M_b=0$  (crossing over to figure 1b), when, in addition to \eqref{v1}, the condition 
\be
v_{pc}^{-2}\int_0^{v_{pc}}v_0f(v_0)dv_0
=\frac{3}{2\pi R^2m}\left(\frac{M}{2GR}\right)^{1/2}\label{v2}
\ee
is satisfied.
We can determine the values of $N$ and $\dot M$ in this case by solving the system of equations (\ref{v1}) and (\ref{v2}).

Let us roughly estimate the ratio of the rates of increase in the mass of DM and BM immediately after the jump from B to D. At point B, these rates are approximately equal and, according to \eqref{e30}, they are inversely proportional to the DM mass density $Nm$. At point D, lying on the upper branch, the value of $\dot M_b$ is the same as at point B. The ratio $\dot M_{DM}$ to $\dot M_b$ at point D is approximately equal to the ratio of $\dot M_{DM}$ at points D and B, which is clearly greater than 1. We can use \eqref{e:27} and set $\dot M_{DM} \simeq C_1 \propto Nm$. So, the ratio of the rates of increase in the mass of DM and BM at point D is proportional to $N^2$ and can be very large at the early stages of galaxy evolution.
\begin{figure}\begin{center}
\includegraphics[width=8cm]{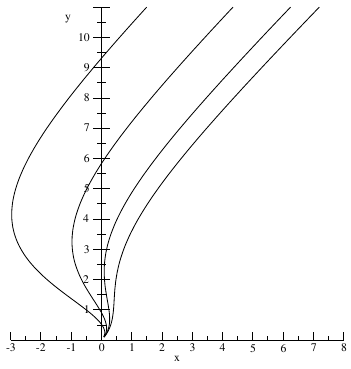}
\end{center}
\caption{\label{f:2}Dependence of the quantities \eqref{xy} proportional to the rates of increase in the total and baryonic masses of the galaxy for the case of the Maxwell distribution of DM particle velocities. The curves from left to right correspond to the different values of the parameter $\mu$ from \eqref{mu} equal to 10, 7, 5 and 4.}
\end{figure}

Let us confirm these arguments with the example of the Maxwell-Boltzmann distribution and use eq. \eqref{e:Max1}. Since we are interested in the ratio of mass gain rates, we introduce the dimensionless variables
\be
x=\gamma \dot M_b,\quad y=\gamma \dot M,\quad\gamma=\sqrt{\frac{2GR}{M}}v_{\max}^{-2}. \label{xy}
\ee
From \eqref{e:Max1} and \eqref{e:Mdotsum} we obtain
\bea
x &=& y-\mu P(y) \label{e:xofy} \\
\mu &=&\frac{2R^2mN}{v_{\max}}\sqrt{\frac{2\pi GR}{M}} \label{mu} \\
 && \hspace{-1cm}\simeq 0.01\frac{mN}{0.3\rho_{c0}/h^2}\frac{100{\rm km}/{\rm s}}{v_{\max}}\left(\frac{R}{{\rm 0.3Mpc}}\right)^{5/2}
 \left(\frac{10^{12} M_\odot}{M}\right)^{1/2} \nonumber
\eea
where the function $P(y)$ is given in \eqref{e:Max2}. In \eqref{e:xofy} the parameter dependence has been reduced to the single dimensionless parameter $\mu$, which decreases with time, mainly due to the decrease in $N$. Figure \ref{f:2} shows four curves corresponding to the values of this parameter equal (from left to right) 10, 7, 5 and 4. They are shown not schematically, as in Fig. \ref{f:1}, but accurately to scale.

At $\mu=4$ the curve is monotonic like in Fig. \ref{f:1}d and this value is slightly below the critical value for the transition to the s-shaped curve. For $\mu=5$ we have a curve similar to that shown in Fig. \ref{f:1}c and this value is slightly below the value of $\mu$ at which the left edge of the top branch intersects the y-axis. These two special values of $\mu$  are rather close, they differ only by a factor 1.25. For larger $\mu$ the curve has the form shown in Fig. \ref{f:1}b. The ratio of the rates of mass gain of DM and BM after the jump to upper branch is approximately 11 at $\mu=5$, 55 at $\mu=7$, and 85 at $\mu=10$. It increases rapidly with increasing $\mu\propto N$. As a result, the galaxy at the stage of intense capture accumulates a lot of dark matter.

\subsection{Qualitative description of catastrophic DM capture}\label{Qd}
After analyzing the conclusions obtained from  eqs.~\eqref{e:Mdotsum} and \eqref{e:Mdmdot} within the framework of our model, we can describe the process of accumulation of dark matter inside the halo of the galaxy.

As a result of the growth of small density fluctuations, regions of increased density emerge, in which galaxies can form. The surrounding matter, both baryonic and dark, begins to fall into them. The infalling BM cools and is captured. In the absence of accretion of baryonic matter, DM particles fly through protogalaxies and are not captured.

The situation changes with an increase in the mass of the baryonic component of the galaxy due to accretion, mergers of galaxies and other processes. A fraction of the DM particles flying into the galaxy at sufficiently low speeds is being captured. In Fig. 1b, this is described by the section AB on the lower stable branch of the s-shaped curve. The mass of dark matter inside the galaxy begins to grow, but the rate of its increase is smaller than the rate of increase in the mass of BM. 

If the rate of increase in the mass of the BM exceeds a certain threshold value, $\dot M_{bc}$, something similar to  a phase transition occurs with a change in the state of the system. In  Fig.~1b, this corresponds to a sharp jump from B to D after reaching the right boundary of the bottom stable branch at point B. The simple estimate \eqref{e30} determines the dependence of the relative 
critical growth rate of baryonic matter $\dot M_{bc}/M$ on the size of the galaxy $R$ and on the mass density of dark matter in the intergalactic space $Nm$ (during the formation of galaxies, this is simply the density of dark matter). The value of $N$ decreases rapidly due to the expansion of the Universe, hence if the jump did not occur during the formation of the galaxy, then it will not occur at later times. The only exception might be the process of merging of galaxies, which can provide a transition to the upper branch due to a sharp temporary increase in the rate of baryonic mass growth, $\dot M_b$.

The critical baryonic mass growth rate required for the jump is smaller for objects with large $R$ (galaxies and their clusters) than for objects with small $R$ (stars, etc.). Therefore, dark halos  form around galaxies, but not around stars.

After a jump to the upper stable branch, the state of the system corresponds to point D or its vicinity. Let us consider the case where the density $N$ is high enough such that $-4b_2/b_1^2<1$ and the $\dot M(\dot M_b)$ curve has the shape 1b. If $\dot M_b$ increases, the system shifts to the right, say, to point E. If  $\dot M_b$ decreases, the system shifts to the left along the curve DG. On this curves, $\dot M_{DM}\gg \dot M_b$. During this phase, galaxies can become DM dominated. Note that in case 1b the capture of DM particles  continues even if the accretion of baryonic matter ceases at point G.

However, not only the value of $\dot M_b$, but also the curve itself changes with time. The dynamics of the change in the curve is associated primarily with the decrease of $N$. The change in the shape of the distribution of the velocities of one DM particle, say, the coefficients $\tilde b_i$, has a much weaker time dependence.

When the threshold value of the intergalactic DM density $N=-4b_2/b_1^2$ is reached, the left boundary of the upper stable branch crosses the y-axis. The curve takes the form shown in Fig. 1c. At point C, the system  can jump to point F on the lower branch. In this case, the mass of the dark halo practically stops growing.

Without knowledge of the DM velocity distribution, we cannot determine the density at which the BD jump occurred, so we do not know whether the curve is described by graph 1b or 1c at a given time. But the transition from curve 1b to curve 1c is inevitable. In the above description, we assume that it happened later than the jump from B to D.

If the state of the system has not descended to the lower branch and intensive capture of DM particles continues due to the high rate of accretion of baryonic matter $\dot M_b$, then with a further decrease in $N$,  the $\dot M(\dot M_b)$ curve becomes monotonic as shown in fig. 1d and represents a single stable branch at densities $N$ below the next threshold value $N_1$ given in eq.~\eqref{N1}. This can be considered the end of the stage of intensive capture of DM particles. It is obvious that this transformation occurs later than the crossing of point C through the $y$-axis.

\section{Some quantitative estimates}\label{Sq}
For an estimation we use the parameters of our Galaxy. The Milky Way cannot be considered typical as there are many more dwarf galaxies in the Universe, but it is a good example of a large galaxy. We set $M\approx10^{12} M_{\odot}\approx 2\cdot 10^{42}$  kg and $R\approx10^6$ ly $\approx10^{22}$m. The last estimate is based on the value $R=292 \pm 61$ kpc \citep{2020MNRAS.496.3929D} and is a rather large value. It is of the order of the average distance between galaxies and slightly less than half the distance to the Andromeda galaxy, M31. For this values  the time of flight through the center of the Galaxy exceeds 2 million years even for an ultra-relativistic particle. But we are more interested in slow particles captured by the Galaxy. As  mentioned above, their initial speed is less than $u=(2GM/R)^{1/2}$, and the speed of passage of the halo is approximately equal to $u=170$ km/s. This gives an upper bound on the time-of-flight of a galaxy $\tau$ for the non-capture case as $\tau\leq4.5\cdot 10^9$ years. 

There are alternative estimation of  $R$. Some of them one can find in the review article by \cite{2013JCAP...07..016N} and in papers by \cite{2014ApJ...794...59K}, \cite{2012PASJ...64...75S}. If we choose the value $R=200$ kpc $\approx 6.5\cdot 10^5$ ly $\approx6.2\cdot 10^{21}$m with the same estimate of $M$, we find $u\approx 210$ km/s, $\tau\leq 10^9$ years. If we choose a lower estimate $R=100$ kpc, then $u\approx 300$ km/s and $\tau\leq3.3\cdot10^8$ years. 

These $\tau$ values are less or much less than the ages of the Universe and of the Galaxy for all estimates of $R$. This confirms the assumption underlying the model that during the passage of a particle that is not captured by the galaxy, the mass of the latter increases, but not by very much. It is clear that this is also true for dwarf galaxies with significantly smaller halo sizes.

We can estimate $v_{pc}$ from \eqref{e:tt}, setting $\dot M=M/T$ with $T\simeq 1.3\times 10^{10}$ years, which corresponds to galaxy formation at $z\simeq5$ to $20$. For the Milky Way we obtain $v_{pc}\approx 100\,$km/s for $R=300\,$kpc, $v_{pc}\approx 60\,$km/s for $R=200\,$kpc, and $v_{pc}\approx 50\,$km/s for $R=100\,$kpc.

We are more interested in estimating the rate of halo mass increase due to DM capture. Let us assume that the Milky Way has not left the stage of intense capture and apply the formula \eqref{e:27}, more precisely, its limit for large $\dot M$. Using the \eqref{e:28} with $\int_0^\infty f(v_0 ) v_0 dv_0 \approx N v_{max}$, we find
\bea
\dot M_{DM}\approx C_1\approx \pi R^2 \rho_{DM}v_{max}\nonumber \\
\approx 0.08 \varkappa h^2\left(\frac{R}{200\, {\rm kpc}}\right)^2 \frac{v_{max}}{100\,{\rm km/s}} M_{\odot} \textrm{ per year.} \label{e:29}
\eea 
Here we have denoted the DM mass density in the intergalactic space as $\rho_{DM}$. In our estimation, we took into account that $Nm=\rho_{DM}= 0.25\varkappa\rho_{c}$. It is less than the average density of dark matter in the Universe, which is now approximately 25\% of the critical density $\rho_{c}$, determined by the Hubble constant $H_0=h100\,$km/s/Mpc. The coefficient $\varkappa<1$ is introduced to account for this difference, which caused by the fact that part of the dark matter is accumulated in the halos of galaxies.

The product  $\dot M_{DM}$ times the age of the galaxy is much smaller than the DM mass in our Galaxy. The reason for this is that the rate of mass increase was significantly higher in the early stages of the capture of dark matter particles by the Galaxy. Let us estimate the mass 
of dark matter $M_{DM}(z_0)$ captured from the time corresponding to the redshift $z_0$ 
to today. We assume that all this time there was an intense capture of particles and the mass gain is described by the equation \eqref{e:29}. The mean density of dark matter in the Universe is proportional to $(1+z)^{3}$. We can set $\varkappa\simeq 1$ for the early stages of galaxy evolution which account for most of the captured DM. 

Let us assume that the galaxy from the beginning of the considered period formed a gravitationally bound system and its dimensions did not increase due to Hubble expansion. It is difficult to estimate by how much $v_{max}$ changes with the expansion of the Universe. On the one hand, the speed of a particle flying far from galaxies and not interacting with other particles remains almost unchanged. On the other hand, an analogy can be drawn with the cooling of an ideal gas during its adiabatic expansion. However, it is doubtful that DM particles would be in a state of thermal equilibrium.

Therefore, and for simplicity, we estimate $M_{DM}(z_0)$, assuming that the values $R$ and $v_{max}$ to be approximately constant during the period under consideration and the change in the capture rate is to be determined mainly by the change in the DM density.

Let us denote the current capture rate as $\dot M_{DM}(0)$ and apply the flat $\Lambda$CDM model. We obtain
\bea
M_{DM}(z_0) &=& \int \dot M_{DM}(0)(1+z)^{3} dt  \nonumber \\
&=&W(z_0)\dot M_{DM}(0)H_0^{-1}\\
\mbox{with} \hspace{0.4cm}&& \nonumber \\
W(z_0) &=&\int_{0}^{z_0} \frac{(1+z)^2}{\sqrt{\Omega_{\Lambda}+\Omega_m(1+z)^3}} dz.
\eea
Here $\Omega_{\Lambda}\approx0.7$ and $\Omega_m\approx0.3$ are density parameters for the cosmological constant and matter. Thus, the mass of dark matter captured by the Galaxy from the moment $z=z_0$ corresponds to that which it would have captured during the time $W(z_0)H_0^{-1}=W(z_0)\cdot 10^{10}/h\,$years if the current rate of capture was maintained.
We can calculate $W(19)\approx114$, $W(24)\approx161$, $W(32.3)\approx250$. We see that, according to our rough estimate, the dark matter that forms the dark halo of the Galaxy can be captured if the process of intense capture begins at $z\simeq20$ or $z\simeq30$.

In order to avoid misunderstanding, we emphasize once again that we do not assert that at the present time the Galaxy continues to actively capture DM particles and its state is now on the upper branch. Almost all DM was captured at the earliest stage of this process. It can be assumed that the process of moving to the lower branch (the fall from C to F in Fig. 1c) has already occurred. We do not know if active capture resumed temporarily during the capture of a single dwarf galaxy (see \cite{2021ApJ...923...92N}). It may also be that some (or most) galaxies never had significant baryonic accretion and never underwent catastrophic DM capture and maintained their ratio of baryonic to dark matter from the initial time of formation.

There are galaxies in the Universe more massive than the Milky Way. When evaluating \eqref{e:29}, we assumed that the capture rate is maximal. Differences in the mass of dark matter in galaxies can be related to the sizes of galaxies and the moment of the beginning and end of intense capture, i.e., the time of the jump to the upper branch and fall back to the lower branch.

The rate of matter capture is proportional to $R^2$. From the estimate \eqref{e30} we can assume that for a larger proto-galaxy the jump to the upper stable branch occurred earlier than for a smaller one. The system can descend to the lower branch not before the  point C crosses the y-axis and the curve takes the form shown in Fig. 1c. But   significant accretion rate of BM allows it to remain on the upper branch for some time after that and to continue to accumulate dark matter at a significant rate. The accretion rate is clearly larger in a big massive galaxy, all other parameters being fixed.
We know galaxies with estimates of $R$ more than $300-400$ kpc. This are e.g. NGC 4889, NGC 4874, ESO 306-17 and others. It can be assumed that their large masses are associated, among other things, with a particularly effective capture of dark matter.

Another possibility is associated with the merger of two galaxies of comparable mass, continuous merging of galaxies in the cluster potential (“galactic cannibalism”), or early merging during cluster formation. An example of such a merger is the giant interacting elliptical galaxy ESO 146-5 (ESO 146-IG 005) in the center of the cluster Abell 3827. Its total mass is $(2.7 \pm 0.4)\times  10^{13} M_\odot$ within $37 h^{-1}$ kpc according to~\cite{2010ApJ...715L.160C}. This estimate was obtained from strong gravitational lensing. The total halo mass of ESO 146-5 is larger. It is perhaps the most massive galaxy in the nearby universe.

In conclusion, if a galaxy or a galaxy cluster is formed from a strong density perturbation and has a larger than average size and a high initial rate of baryonic mass increase $\dot M_b$, it will accumulate more DM.

\section{Conclusions}\label{s:con}
	We studied the  capture of DM particles passing through a galaxy. It is associated with an increase in the mass of the galaxy, primarily its dark halo. The kinetic energy of a particle, which increases as it enters the halo and decreases while it leaves it again, may become insufficient for the particle to leave the halo if the galaxy mass increases sufficiently during the passage. This requires the combined action of two factors. One is an increase in the mass of the baryonic component of the galaxy, and the other is determined by the particle flux. Both have to be sufficiently large for significant capture of DM particles. Furthermore, both change in time, leading to quick changes in the DM capture rate. A capture occurs precisely by objects of the size of galaxies, but not by much smaller astronomical objects like stars.

As a result, the particle is captured and begins to move inside the gravitational potential well of the galaxy. The capture process can be described by catastrophe theory. DM accretion can jump from a moderate capture rate of order the baryonic mass growth to a much large value which we denote {\it catastrophic DM capture}. Its start may even be as early as the non-linear growth of primordial density fluctuations during the Dark Ages. The ratio of the influx rates of dark and baryonic matter can be very significant during catastrophic DM capture which may explain the large observed DM to BM ratio in certain galaxies.

The growth rate of the mass of baryonic matter inside a galaxy, for example due to accretion and cooling or due to galaxy cannibalism, must exceed a certain threshold value to enter the catastrophic capture regime. Also, the density of DM particles in intergalactic space must exceed a certain threshold value in the  catastrophic DM capture mode. Taking into account that the matter density decreases with time both due to the Hubble expansion and because of the capture of particles as discussed in this paper, it can be assumed that the capture process has either weakened significantly in the past, or will do so in the future.

Particles with  sufficiently high initial velocity can fly through the galaxy, leaving it with a reduced speed due to the action of the mechanism under consideration. The higher the initial velocity, the smaller the loss of both velocity and energy of the particle. As a result, a general decrease in energy and a change in the velocity distribution of the particles occur. This process is more efficient if the galaxy is in a cluster rather than in a void.

A qualitative description of the formation of a dark halo around galaxies is given in Section \ref{Qd}. The process includes several transformations and changes in the state of the  system. Particularly strong fluctuations lead to the appearance of large galaxies, often in clusters. Their size and high mass accretion rates ensure the capture of almost all DM particles that enter inside. 
Some quantitative estimates of the considered process in the Milky Way galaxy are presented in Section \ref{Sq}.

We believe that the approach proposed in this work, in particular the idea of a sharp transition to a regime of intense DM particle capture, can supplement our understanding of the formation of the dark halos of galaxies.  Note also, that the velocity dispersion of DM is neglected in the initial conditions of N-body simulations where it is assumed that DM particle velocities are fixed exactly by the peculiar velocity field. Even if DM is expected to be cold so that velocity dispersion is probably small, our effect might help to lead to earlier galaxy formation and explain the surprising data JWST~\cite{2022arXiv220712446L,2022arXiv220805473F}.

The work presented here is preliminary as we just show the main features studying a toy model. {Within this toy model we can demonstrate the existence of a mode of intense DM particles capture with a catastrophic transition to this mode and back, focusing on the physical aspects of the process. On the other hand, the picture described in the article is certainly simplified. For simplicity we assume spherical symmetry of the halo and we neglect peculiar motion in a reference frame in which the distribution of DM particle velocities is isotropic. 
In addition we implicitly consider the capture of particles by an already sufficiently formed galaxy. However, during the formation of a galaxy from the initial over-density, the density contrast and the DM accretion rate increase from small, linear initial conditions. For a more adequate treatment, a more detailed model will be needed, which considers the capture of dark matter during the growth of density fluctuation at its different stages, including nonlinear growth. A more realistic N-body simulation, including hydrodynamical effects of baryons}  is  needed to show that catastrophic DM capture may be truly relevant for cosmological structure formation.
Another possible continuation of this work is related to the statistical behaviour of a system which is not in thermodynamic equilibrium. DM particles after passing through a growing galaxy are slowed down, lose part of their speed. It would be interesting to investigate the influence of this process on the particle velocity distribution by writing a Boltzmann transport equation for this process.
\vspace{1cm}

{\bf Acknowledgement}
RD acknowledges support from the Swiss National Science Foundation.
SP expresses gratitude to the people and the government of the Swiss Confederation for supporting Ukrainian scientists in wartime. He thanks SwissMAP for funding his visit to the University of Geneva in Spring 2022, and the Département de Physique Théorique  and CERN for the opportunity to prolong  this visit.

\bibliography{refs}

\end{document}